\documentclass[preprint,amsmath,amssymb,aps]{revtex4-1}
\usepackage{graphicx}

\newcommand{\barT}{T}
\newcommand{\WTA}{\textit{WTA}}

\begin{document}

\title{Fermionic Networks: \\Modeling Adaptive Complex Networks with Fermionic Gases}
\author{Marco Alberto Javarone}
\email{marcojavarone@gmail.com}
\affiliation{
Dept. of Mathematics and Computer Science, University of Cagliari, 09123 Cagliari, Italy\\
DUMAS - Dept. of Humanities and Social Sciences, 07100 Sassari, Italy
}

\date{\today}

\begin{abstract}

We study the structure of Fermionic networks, i.e., a model of networks based on the behavior of fermionic gases, and we analyze dynamical processes over them.
In this model, particle dynamics have been mapped to the domain of networks, hence a parameter representing the temperature controls the evolution of the system.
In doing so, it is possible to generate adaptive networks, i.e., networks whose structure varies over time. %
As shown in previous works, networks generated by quantum statistics can undergo critical phenomena as phase transitions and, moreover, they can be considered as thermodynamic systems.
In this study, we analyze Fermionic networks and opinion dynamics processes over them, framing this network model as a computational model useful to represent complex and adaptive systems.
Results highlight that a strong relation holds between the gas temperature and the structure of the achieved networks. Notably, both the degree distribution and the assortativity vary as the temperature varies, hence we can state that fermionic networks behave as adaptive networks. 
On the other hand, it is worth to highlight that we did not find relation between outcomes of opinion dynamics processes and the gas temperature. 
Therefore, although the latter plays a fundamental role in gas dynamics, on the network domain its importance is related only to structural properties of fermionic networks.
\end{abstract}
\maketitle
\section{Introduction}
Nowadays, modern network theory~\cite{albert01,estrada01} represents a growing research field characterized by a strong interdisciplinarity as several systems can be mapped onto complex networks, e.g., the World Wide Web, social networks, brain networks, and financial networks~\cite{estrada02,barabasi01,capocci01,sporns01}.
Several models to achieve networks provided with specific features, e.g., homogeneous and heterogeneous topologies, small-world behaviors, and multi-level structures~\cite{boccaletti01}, have been proposed in recent years.
Among the most well studied network models, we recall the Barabasi-Albert model~\cite{barabasi02}, the Erd\"{o}s-Renyi graphs~\cite{erdos01} and the Watts-Strogatz model~\cite{watts01}.
The Barabasi-Albert model (BA model hereinafter), based on the preferential attachment mechanism, allows one to generate scale-free networks. This kind of network has a degree distribution that follows a power-law~\cite{albert01}.
The Erd\"{o}s-Renyi graphs model generates `'classical random networks'', i.e., networks provided with a binomial degree distribution, that converges to a Poissonian distribution under opportune conditions.
Finally, the Watts-Strogatz model, based on an interpolation between classical random networks and regular ring lattices, allows one to achieve networks characterized by a small-world behavior.
In principle, a network can be viewed as a dynamical system that evolves over time, as new nodes or new links can be added to the system in every moment; for instance, the number of Facebook users varies with hourly frequency, companies draw up new relations with other companies or customers, and so on.
Therefore, in order to represent real networks, often it is important to use adaptive networks~\cite{gross01}, characterized by a structure that varies over time. 
Adaptive networks can be generated in different ways, in general starting from a topology and introducing a stochastic rule to let the network vary over time. Furthermore, some models of adaptive networks have been inspired from theoretical physics. In particular, the bosonic networks~\cite{bianconi01,bianconi02} and fermionic networks~\cite{bianconi03,javarone01} represent two models inspired from the dynamics of quantum gases. Moreover, both quantum models show that network evolution can be represented in terms of phase transitions.
Therefore, under this perspective, complex networks are considered as thermodynamic systems that evolve over time~\cite{hartonen01}.
In this work we focus on the model of fermionic networks~\cite{javarone01}, achieved by mapping complex networks to fermionic gases. In particular, we analyze both the structure and the outcomes of a dynamical process.
The aim is to show how fermionic networks can be used as a computational framework to generate adaptive networks and to study complex systems.
Eventually, it is worth recalling that a number of models `inspired from'/'based on' quantum mechanics have been developed in the field of complex networks, opening the way to the emerging field of 'quantum complex networks' --see~\cite{zimboras01,biamonte03}.  
The remainder of the paper is organized as follows: Section~\ref{sec:networks-models} describes the model of bosonic networks and that of fermionic networks. Section~\ref{sec:structure} shows the results of analysis performed to investigate the structure of fermionic networks. Section~\ref{sec:processes} is devoted to illustrate results of opinion dynamics over fermionic networks, using as reference the classical voter model. We conclude in Section~\ref{sec:conclusions}. 

\section{Quantum Complex Networks}\label{sec:networks-models}
Now, we will briefly recall the model of bosonic networks~\cite{bianconi01}, and later illustrate more deeply the fermionic network model~\cite{javarone01}. In particular, the latter constitutes the model whose behavior is investigated in this work. 
From a computational perspective, these two 'dual' models allow one to represent different phases of an evolving network, e.g., from classical random to scale-free configurations. Therefore, in principle they can be considered as models to generate adaptive networks. 
\subsection{Bosonic Networks} 
The model of bosonic networks has been developed by Bianconi and Barabasi~\cite{bianconi01}. It allows one to compare a network evolution to a phase transition of bosonic gases, as two main structures (i.e., \textit{fit-get-rich} and \WTA) are identified as two different phases at low temperatures. In bosonic networks, each node represents an energy level and each link a pair of particles. In doing so, it is possible to perform the mapping between the two domains, i.e., from quantum gases to networks and vice versa. 
Moreover, a fitness parameter $\eta$ is introduced in order to compute the energy:
\begin{equation} \label{eq:energy_fitness}
\epsilon = -\frac{1}{\beta} \cdot \log \eta
\end{equation}
with $\beta = \frac{1}{T}$. In this context, the fitness $\eta$ describes the ability of nodes to compete for new links. In particular, the $i$th node has a probability to connect with new nodes proportional to:
\begin{equation} \label{eq:prob_link}
\Pi_{i} = \frac{\eta_{i} k_{i}}{\sum_{j} \eta_{j}k_{j}}
\end{equation}
with $k_{i}$ degree (i.e., the number of neighbors) of the $i$th node. Hence, new nodes tend to generate connections with pre-existing nodes having high values of $(\eta,k)$. 
Scale-free networks in the \textit{fit-get-rich} phase are characterized by a power-law equation (later illustrated more deeply), and they have a small fraction of nodes with a high degree (i.e., value of $k$) connected to many others with a low degree. 
Particles of a bosonic gas occupy lower energy levels when the temperature decreases. Then, Bose-Einstein condensation takes place at a temperature below the critical one $T_{c}$. 
In bosonic networks, as the temperature decreases, some particles move to lower levels while keeping the corresponding ones at upper levels (recall that each link is mapped to two particles). In doing so, links concentrate on a few nodes, until they form a condensate in the \WTA~phase. This is characterized by the fact that only one node dominates.
Eventually, in~\cite{bianconi02} quantum statistics of bosonic networks is more deeply investigated.
\subsection{Fermionic Networks}
\begin{figure}[!ht]
\centering\includegraphics[width=5.0in]{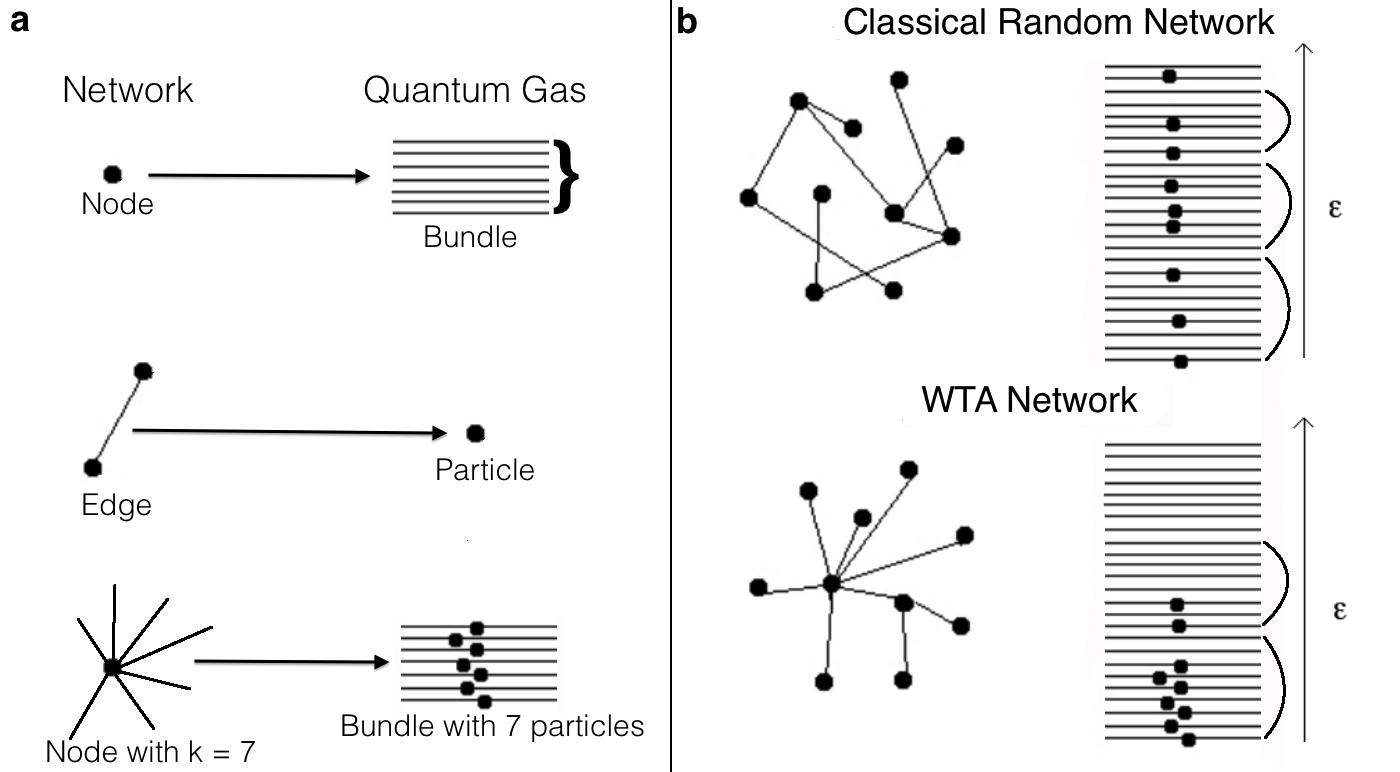}
\caption{\label{fermionic_network_scale} Mapping networks to quantum gases. \textbf{a}) From top to bottom: one node mapped to a bundle, one edge (between two nodes) mapped to a particle, and one node with a degree $k = 7$ (i.e., with $7$ edges) mapped to a bundle with $7$ particles. \textbf{b}) On the left, from top to bottom, the evolution of a network with $10$ nodes and $9$ links from a classical random network to a \WTA~network. On the right, their corresponding fermionic models, showing most deep bundles (by drawing simple arches), which result from a cooling process that pushes particles to low energy levels.}
\end{figure}

The first attempt to model networks as fermionic gases has been proposed~\cite{bianconi03}, where the author represents growing dynamics of a Cayley tree by the Fermi distribution.
As for bosonic networks and for Cayley trees described by the Fermi distribution, the fermionic network model proposed in~\cite{javarone01} has been inspired from the behavior of quantum gases.
It is worth to highlight that fermionic models, described in~\cite{bianconi03} and in~\cite{javarone01}, have been developed by a different mapping between the quantum world and the networks world. As a consequence, these two models lead to very different structure of networks.
It is also important to highlight that, as we stated above, the mapping task performed to define the model proposed in~\cite{javarone01} followed a statistical approach inspired by the fermionic distribution, i.e., the fermionic behavior is mapped only on a quality level. 
As result, although further analytical investigations are still required in order to achieve a full comprehension about the dynamics of the proposed model~\cite{javarone01}, the latter allows, from a computational perspective, to define a framework to study adaptive networks.
Hereinafter we refers to~\cite{javarone01} when discussing about fermionic networks, that are now briefly introduced.
Quantum gases assume different configurations, in terms of particles distribution among energy levels, depending on their temperature. In particular, although fermionic gases have a particles distribution that follows the Fermi-Dirac statistics, in the high-temperature limit, they show a quantum-classical transition. The latter implies their particles distribution is approximated by the Maxwell-Boltzmann distribution at high temperature.
Moreover, at low temperatures (i.e., as $T \to 0$) particles move to lower energy levels until they occupy the deeper bundles, with only one particle per energy level (due to the Pauli exclusion principle).
Since, in the proposed model, the concept of bundle of energy levels has a central role, we briefly recall its physical meaning. Usually, quantum energy levels are very closely spaced, and their amount is much greater then the amount of particles. In these systems, a bundle represents a group of energy levels having, approximately, the same energy.
In order to introduce the fermionic network model, it is important to focus on the different configurations that a fermionic gas assumes varying the temperature.
Now, let us consider a simple network, where nodes are mapped to bundles and edges (or links) are mapped to particles of a fermionic gas ---see panel \textbf{a} of Figure~\ref{fermionic_network_scale}.
Usually, in these gases the number $g_{i}$ of available states in the $i$th bundle is much bigger than the amount $p_{i}$ of its particles.
The energy $\epsilon_{i}$ of the $i$th bundle can be randomly assigned or can be computed in accordance with a property of the system, e.g., a fitness parameter $\eta$ as for the bosonic networks.
It is worth highlighting that, in this model, lower bundles have more energy levels. In particular, the first bundle has $n-1$ levels, the second has $n-2$ levels, and so on.
In doing so, the link $l_{ij}$, between the $i$th and $j$th nodes, is represented only by a single energy level (i.e., $\epsilon_{ij}$), which in turn belongs to the $i$th bundle. In this last example, we assume that the $i$th bundle is deeper than the $j$th.
Therefore, the last node is represented by a bundle without energy levels. Notwithstanding, the last node (e.g., $y_0$) can be linked to another node (e.g., $x$) if a particle stays at the $\epsilon_{xy_0}$ level.
Remarkably, low energy bundles represent nodes that, at low temperature, become hubs, i.e., nodes with a high degree. On the other hand, high energy bundles represent nodes with few neighbors (i.e., with low degree) at low temperatures. 
Since fermionic networks aim to behave as fermionic gases, their dynamics have a fundamental role.
In particular, in this context we deal with adaptive networks, i.e., networks that vary over time as many real networks do. 
In order to make the model as simple as possible, we consider fermionic networks as closed systems, hence the number of nodes and the number of links are constant.
The energy of each bundle (i.e., node) can be computed by Eq.~(\ref{eq:energy_fitness}), then the relative position of each bundle depends on the value of its energy and, as discussed before, deeper bundles embody more states. 
Particles are able to jump among energy levels as the temperature varies, therefore at high temperatures particles follow the classical Maxwell-Boltzmann distribution.
Instead, as the temperature decreases, particles move to lower energy levels (see panel \textbf{b} of Figure~\ref{fermionic_network_scale} ---from~\cite{javarone01}).

The dynamics of fermionic networks depend on the system's temperature $T$, hence we can analyze the outcomes of the model in terms of network evolution varying the value of $T$, i.e., by cooling and by heating processes.
\\\\
\textbf{\textit{Cooling process.}} During a cooling process, only a few nodes gain new links and their degree $k_{i}$ increases.
In the event the number of particles approximates that of bundles, as temperature decreases the \WTA~phase takes place \textemdash see also \cite{bianconi01}. 
Every time the temperature varies, each particle changes position (i.e., bundle) with a probability computed as:
\begin{equation} \label{eq:jump_probability}
p(i \to j) = \frac{\Delta T}{\barT} \cdot \frac{1}{\Delta B(j,i)} \cdot  f(g_{j})
\end{equation}
\noindent  where $i$ and $j$ are the starting and the ending bundle, respectively, $\barT$ is the system temperature before the variation, $\Delta T$ the variation of temperature, $\Delta B(j,i)$ the distance between the bundles $j$ and $i$, and $f(g_{j})$ the function:
\begin{equation} \label{eq:free_space}
f(g_{j}) = 
\begin{cases} 0 & \mbox{if  $g_{j} = 0$} \\ 
                                  1 & \mbox{if  $g_{j} \ge 1$}
\end{cases}
\end{equation}
with $g_{j}$ number of available states in the $j$th bundle.
In particular, it is worth to highlight that the distance between bundles (i.e., $\Delta B(j,i)$) is computed by considering their positions in the energy scale. For instance, the distance between the bundle $B_0$ (i.e., lowest bundle) and the bundle ($B_3$) (i.e., fourth bundle, identified starting from the lowest level) is $\Delta B(3,0) = 3$, since a particle has to perform three jumps to reach $B_3$ starting from $B_0$ (or vice versa).
Since a particle in the $i$th bundle can jump to underlying bundles (as defined in Eq.~(\ref{eq:jump_probability})), the probability $p_J$ to jump from the $i$th to another bundle is computed as follows:
\begin{equation} \label{eq:total_jump_probability}
p_{J}(i) = \sum_{z=1}^{i-1} p(i \to z)
\end{equation}
hence, the probability $p_S$ to stay in the same bundle is:
\begin{equation} \label{eq:stay_probability}
p_{S}(i) = 1 - p_{J}(i)
\end{equation}
In so doing, the final bundle of each particle is chosen by a weighted random selection among all possible bundles (including that one in which the particle is located). It is worth to mention that Eq.~(\ref{eq:jump_probability}) has been properly devised, in order to perform a mapping from the quantum system to the network structure. Moreover, although Eq.~(\ref{eq:total_jump_probability}) embodies a harmonic series (i.e., the distance between bundles), the ratio $\frac{\Delta T}{T}$ and the function defined in Eq.~(\ref{eq:free_space}) allow to ensure the final convergence to values smaller than or equal to $1$. Anyway, as for other parts of the proposed model (before discussed), further investigations are required to define analytical laws more relevant in the physical context of quantum gases. 
\\\\
\textbf{\textit{Heating process.}} On the other hand, during a heating process particles move to higher energy levels. Now, for every variation of the temperature, the probability for a particle to change bundle (e.g., from the $i$th to the $j$th) is computed using a variant of Eq.~(\ref{eq:jump_probability}); in particular, the function $f(g_{j})$ is defined as:

\begin{equation} \label{eq:free_space_heat}
f(g_{j}) = 
\begin{cases}
0 & \mbox{if  $g_{j} = 0$} \\ 
                                  1 - \frac{p_{j}}{g_{j}} & \mbox{if  $g_{j} \ge 1$}
\end{cases}
\end{equation}
\noindent with $p_j$ number of particles in the $j$th bundle. 
Eq.~(\ref{eq:free_space_heat}) has been devised to avoid that, at high temperatures, particles fill densely a few high energy levels.
Hence, each particle changes position with probability:
\begin{equation} \label{eq:total_jump_probability_heat}
p_{J}(i) = \sum_{z=i + 1}^{n-1} p(i \to z)
\end{equation}
whereas, each particle stays in its bundle (i.e., it does not jump) with probability defined by Eq.~(\ref{eq:stay_probability}).
As before, a weighted random selection is performed for choosing the final position of each particle.
In this model, the temperature represents a parameter leading to an evolution of the system. Therefore, when fermionic networks are used to study some other models, it is worth to properly map the temperature to a relevant parameter of the system, i.e., a parameter that drives its evolution.

\section{Fermionic Networks: Structure}\label{sec:structure}
In this section, we show some structural properties of fermionic networks, i.e., the degree distribution, the assortativity and the clustering coefficient, computed during the evolution of the system. We recall that the evolution is performed by varying the temperature and, moreover, we highlight that the temperature-step adopted in each simulation is equal to $1$.
A short description of each listed property is provided before to show the related results.
Eventually, we generated fermionic networks by two methodologies: 
\begin{enumerate}
\item Starting from an E-R graph $(N,\zeta)$ (with $N$ nodes and $\zeta$ probability of each link to exist), and then mapping the behavior of the related gas to the network as the temperature varies;
\item Starting from the gas, adding particles to each energy level with probability $\zeta$, and then generating the related graph from the particle distribution, mapping the behavior of the gas to the network as the temperature varies.
\end{enumerate}
We highlight that, by both methodologies, we start with an E-R graph at $t=0$, before decreasing the temperature and later increasing it. 
Obviously, other different methodologies can be used to perform this task, e.g., starting with a scale-free network or by spreading particles among energy levels with different distributions.
\subsection{Degree Distribution}
The degree distribution, defined $P(k)$, represents the probability that a randomly chosen node had the degree (i.e., the number of neighbors) equal to $k$.
Probably, the degree distribution is the most important characteristic in order to investigate the structure of a network. For instance, it allows to know if a network has hubs (i.e., nodes with a high degree) and if it is homogeneous or not.
Moreover, we recall that two famous models of random networks before cited (i.e., E-R graphs~\cite{erdos01} and scale-free networks~\cite{barabasi02}) are easily identified by their degree distribution. E-R graphs have a Poissonian distribution, whereas scale-free networks have a distribution that follows a power-law.
In particular, E-R graphs have a $P(k)$ defined as follows:
\begin{equation} \label{eq:poisson}
P(k) \sim e^{-\zeta n} \cdot \frac{(\zeta n)^{k}}{k!}
\end{equation} 

\noindent with $n$ number of nodes and $\zeta$ probability of each edge to exist.
On the other hand, scale-free networks have a $P(k)$ that follows the equation:
\begin{equation} \label{eq:scale_free}
P(k) \sim c \cdot k^{-\gamma}
\end{equation} 
\noindent with $c$ normalizing constant and $\gamma$ scaling parameter (usually in the range $[2,3]$). One of the main differences between these two topologies is that E-R graphs are homogeneous networks, whereas scale-free networks are heterogeneous.
The degree distribution of fermionic networks, computed varying the system temperature, has already been studied in~\cite{javarone01}. 
Here, we show results related to a cooling process in order to illustrate how the network structure is strongly affected by the variation of temperature and, moreover, to show that a transition between E-R graphs to scale-free networks can be achieved by this model.
\begin{figure}[!ht]
\centering
\centering\includegraphics[width=5.0in]{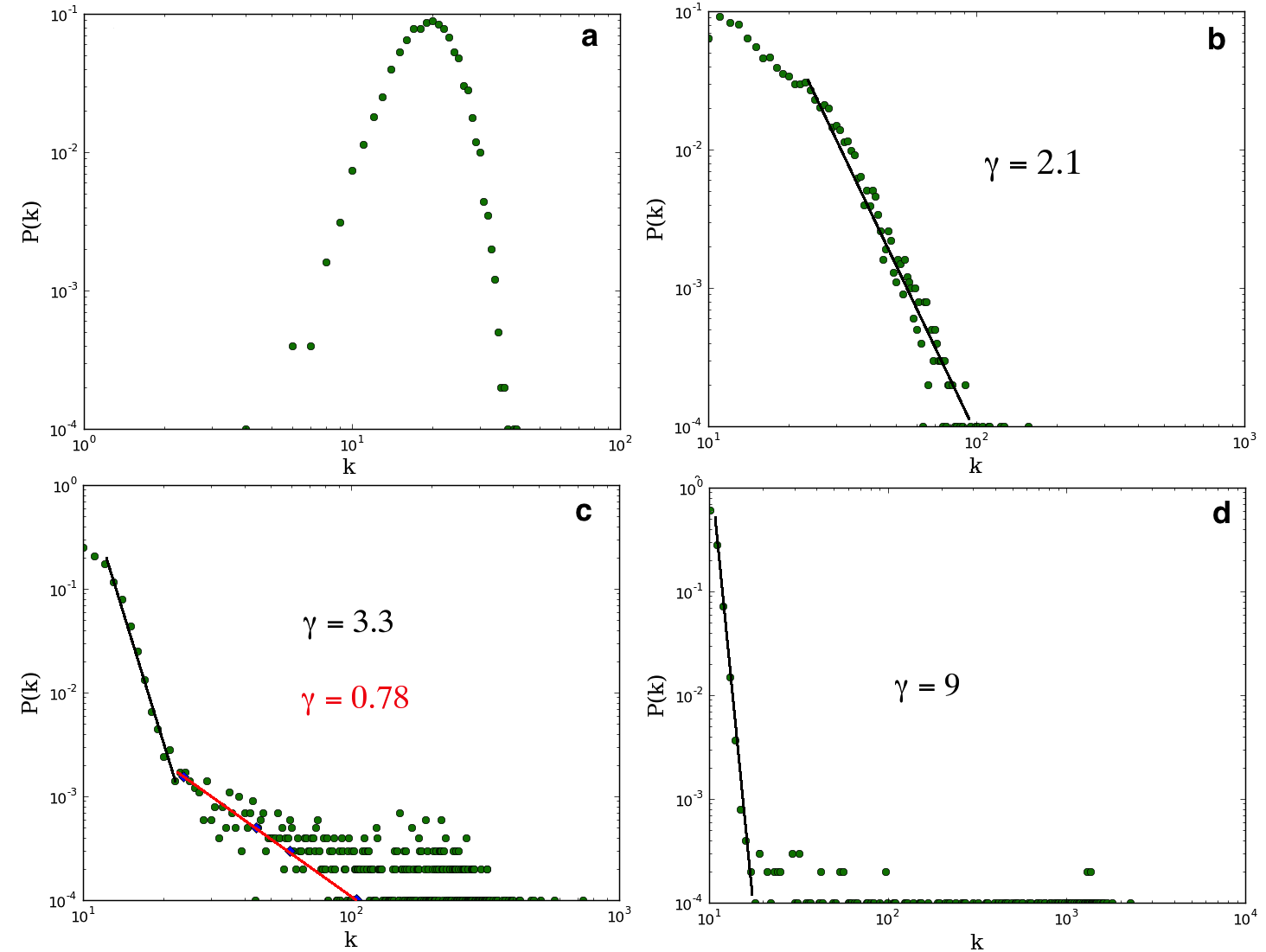}
\caption{The evolution of the $P(k)$ of fermionic networks, during a cooling process, with $10000$ nodes. Each panel shows the fermionic network at a different time step $t$: \textit{a}) at t=0; \textit{b}) at $t=4$;\textit{c}) at $t = 19$; \textit{d}) at $t = 50$. Note that at $t = 0$ the fermionic network has an E –R graph structure, whereas for $t = 50$ it has a WTA structure. Continuous black and red lines are used to highlight data interpolation. The corresponding scaling parameter(s) $\gamma$ is (are) indicated in each panel.}%
\label{fermionic_network_degree}
\end{figure}
As shown in Figure~\ref{fermionic_network_degree} (from~\cite{javarone01}), starting from a E-R structure and decreasing the temperature $T$, the fermionic network achieves a scale-free configuration only after a few steps. Furthermore, other degree distributions characterized by more than one scaling parameter $\gamma$ (identified by a binning process) emerge. Finally, at low temperatures the degree distribution is exponential and, forcing a power-law behavior, we computed a value of $\gamma \sim 9$.
\subsection{Assortativity}
\begin{figure}[!ht]
\centering
\centering\includegraphics[width=5.0in]{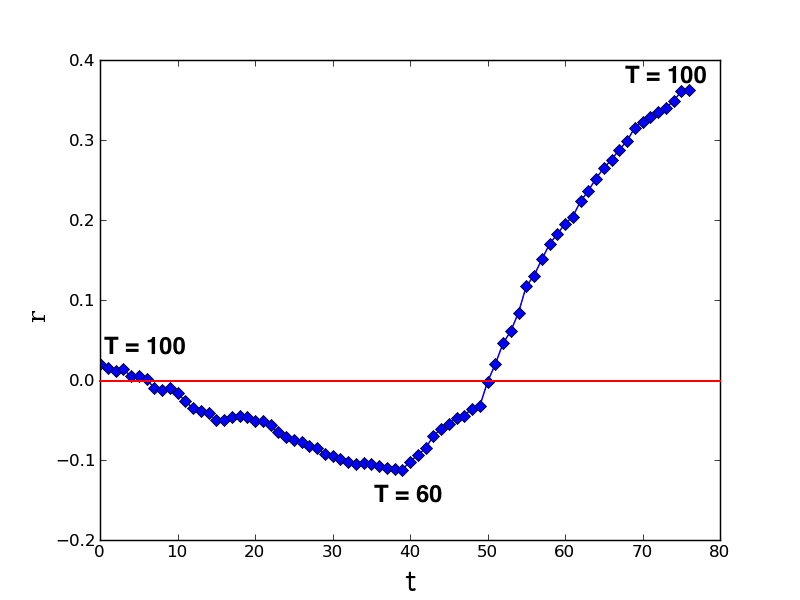}
\caption{Assortativity $r$ of fermionic networks (with $2000$ nodes) over time, varying the temperature of the system. Critical points (maximum and minimum values) are labeled with the related temperature. The red line highlights the value $r = 0$ (i.e., zero assortativity).}
\label{fermionic_network_assortativity}
\end{figure}

Assortativity is a relevant property of networks that allows to evaluate to which extent nodes prefer to attach to other nodes that are (not) similar~\cite{newman01}. In general, networks can be assortative or disassortative, i.e., nodes prefer to attach to those that are similar or different and, as shown in~\cite{torres01}, scale-free networks tend to be disassortative due to an entropic underlying principle. 
Assortativity, indicated as $r$, can be computed by considering the quantity $e_{ij}$, corresponds to the fraction of edges in a network that connects a node of type $i$ to one of type $j$. In accordance with this view, the value of $r$ is calculated as:
\begin{equation}\label{eq:assortativity}
r = \frac{\sum_{i} e_{ii} - \sum_{i} a_{i}b_{i}}{1 - \sum_{i} a_{i}b_{i}}
\end{equation}
\noindent with $a_{i} = \sum_{j} e_{ij}$ and $b_{j} = \sum_{i} e_{ij}$.
A network is assortative when $r$ is positive and, on the contrary, it is disassortative when $r$ is negative.
It is worth to highlight that the similarity can refer to several properties of nodes, as for instance their degree $k$. 
We analyzed the assortativity of fermionic networks varying the temperature from $T=100$ to $T=60$ and vice versa. Figure~\ref{fermionic_network_assortativity} shows the related results.
It is important to note that, as shown in Figure~\ref{fermionic_network_assortativity} , we let the temperature varies at each time step. Therefore, at time $t=0$ $T=100$, at time $t=40$ $T=60$, and finally at $t=79$ $T=100$.
It is interesting to note that, starting with a slightly assortative network at $t=0$, a cooling process let the value of $r$ decreases, then it increases to values higher then zero (i.e., assortative networks emerge).
As the cooling process lead the network to a scale-free configuration in a few time steps (see Figure~\ref{fermionic_network_degree}, this result is in full accordance with the phenomena described by~\cite{torres01}, as scale-free networks have a higher probability to be disassortative than assortative.
Later on, as the temperature increases exponential structures, which describe homogeneous networks, emerge and the mixing degree turn to assortative.
It is worth to highlight that although the assortativity varies as the temperature varies, its value is related to the nature of the achieved network and not directly to the value of the temperature. For instance, as the temperature variation generates a scale-free like network, the expected assortativity is negative (see ~\cite{torres01}), no matter the value of the temperature.
\subsection{Clustering Coefficient}
\begin{figure}[!h]
\centering
\centering\includegraphics[width=5.0in]{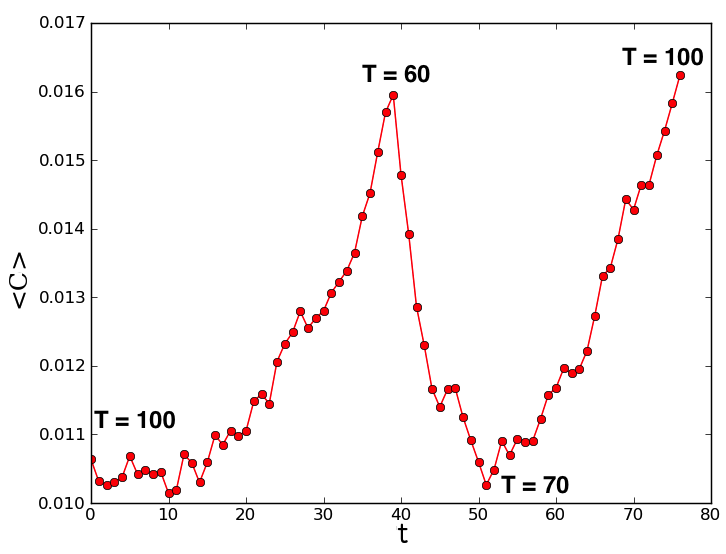}
\caption{Clustering coefficient of the whole fermionic networks (with $2000$ nodes) over time, varying the temperature of the system. Critical points (maximum and minimum values) are labeled with the related temperature.}
\label{fermionic_network_clustering}
\end{figure}

The clustering coefficient allows to measure to which extent nodes in a network cluster together. For instance, in social networks it is possible to identify groups of people where every person knows all the others. 
This property, usually indicated as $C$, has a value in the range $0 \le C \le 1$.
There are different methods to compute $C$ as that defined by Watts and Strogatz in~\cite{watts01}; in particular, they compute local values of the clustering coefficient (for each node) as follows:
\begin{equation}\label{eq:avg_cluster_watts_local}
C_{i} = \frac{Tn_{i}}{Tp_{i}}
\end{equation}
\noindent with $Tn_{i}$ number of triangles connected to node $i$ and $Tp_{i}$ number of triples centered on node $i$. 
In the event the degree of a node is $0$ or $1$, its value of $C$ is $0$. Then, it is possible to compute the clustering coefficient of the whole network as:
\begin{equation}\label{eq:avg_cluster_watts_global}
<C> = \frac{1}{n} \sum_{i} C_{i}
\end{equation}
We analyzed the clustering coefficient in fermionic networks when the temperature varies from $T=100$ to $T=60$ and vice versa --see Figure~\ref{fermionic_network_clustering}.
As before, the temperature varies of $\pm 1$ degree (in accordance with an increasing/decreasing process) at each time step.
It is worth to observe that the variation over time of $<C>$ shows three critical points. In general, variations of temperature in both directions seem to increase the clustering coefficient; with the exception of the time steps which correspond to the inversion of the process, from cooling to heating. In particular, as the system is heated from $T = 60$ to $T= 70$ after a previous cooling process, the clustering coefficient rapidly decreases.
This behavior does not seem related to the structure of generated networks therefore, in principle, the causes should be further investigated. At the same time, it is important to note that the variation of the clustering coefficient is very small if compared to the whole range that $<C>$ can take. Therefore, the phenomenon observed between $T=60$ and $T=100$, as the other increasing tendency, can be considered as casual fluctuation around a value of $<C> \in [0.01, 0.016]$. 
Eventually, we highlight that the computed range of $<C>$ is very small compared to that of small-world networks (characterized by higher values of the clustering coefficient, in full accordance with theoretical expectation for these classes of networks (i.e., E-R graphs and scale-free networks and the intermediate phases).
\section{Fermionic Networks: Opinion Dynamics}\label{sec:processes}
\begin{figure}[!ht]
\centering
\centering\includegraphics[width=5.0in]{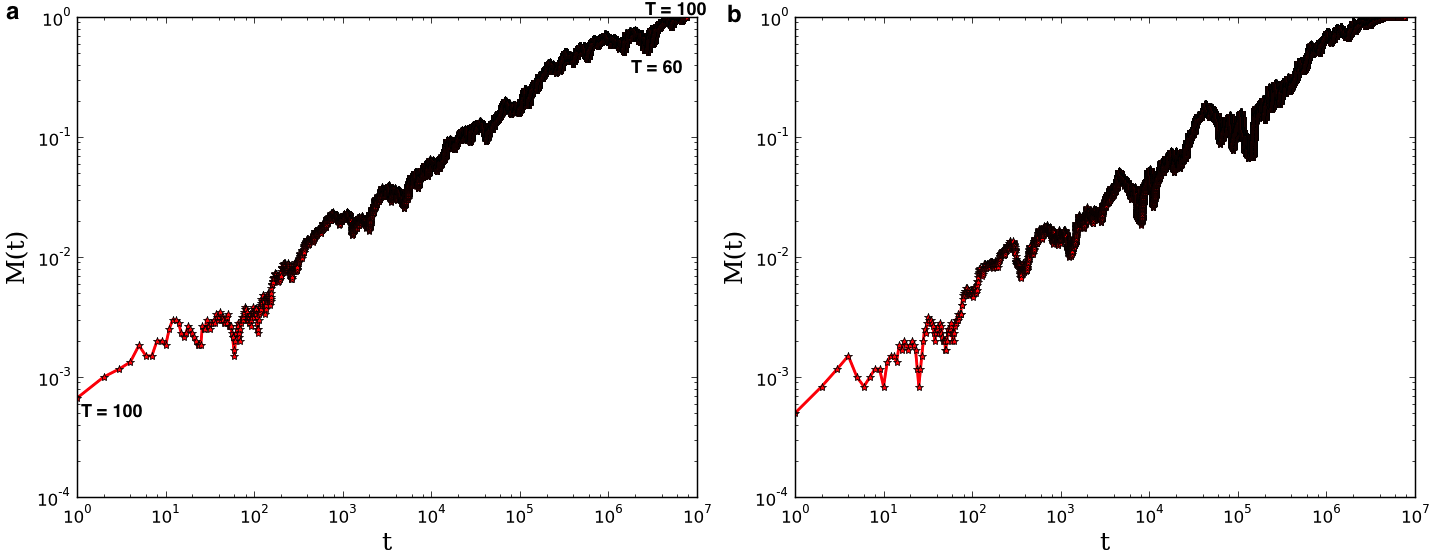}
\caption{Magnetization of the system, with $2000$ nodes, over time. \textbf{a} Results achieved in fermionic networks during a cooling and a heating process. Temperatures $T = 100$ and $T = 60$ are qualitatively indicated, as in the log-scale it is difficult to identify the single points. \textbf{b} Results achieved in classical random networks.}
\label{fermionic_network_voter}
\end{figure}

Here, we study dynamical processes on fermionic networks, focusing our attention on those related to opinion dynamics~\cite{loreto01}.
During last years, opinion dynamics has attracted the attention of several scientists and many models, to study the generation and the spreading of opinions, have been developed (e.g., ~\cite{galam01,galam02,krapviski01,sznajd01,javarone02,javarone03,oliveira01}). 
In these dynamics, both the interactions among individuals and the structure of their network have a fundamental role~\cite{loreto01,miguel01,barrat01}. 
The most simple and famous model of opinion dynamics is the voter model~\cite{redner01,galam02,galam03}, and it can be easily implemented on networks with different topologies.
In this model, at each time step a randomly chosen agent define its opinion in accordance with that of one of its neighbors (randomly chosen). 
In particular, the model considers a set of agents that change opinion over time, by interacting among themselves. 
In particular, the voter model is composed by the following steps: 
\begin{enumerate}
\item randomly select an agent in the population; 
\item the selected agent takes the opinion of one of its neighbors (always randomly selected)
\item repeat from $(1)$ until all agents share the same opinion (or the maximum number of time steps is elapsed)  
\end{enumerate}
Usually, these simple steps are repeated until the global consensus is reached, i.e., an ordered phase emerges in the population.
The voter model allows one to represent the evolution of a population toward consensus in the presence of different opinions. In general, from a physical perspective, by this model it is possible to represent phase transitions from a disordered state to an ordered one of a system~\cite{mobilia01,vespignani01}; although, as shown in~\cite{schweitzer01}, it is possible to introduce a non-linear dynamics that entails the system reaches a final phase characterized by the coexistence of different opinions.
In this model, opinions are mapped to states (or spins), e.g., $\sigma = \pm 1$ or $\sigma \in{0,1}$. In doing so, a relevant parameter that can be analyzed is the magnetization of the system defined as follows:
\begin{equation}\label{eq:magnetization}
M = \frac{|S_0 - S_1|}{N}
\end{equation}
\noindent with $S_0$ and $S_1$ summations of agents in the state $0$ and $1$ (or $-1$ and $+1$), respectively\textemdash see also \cite{javarone03}.
The topology of the agent network plays a relevant role in these dynamics, and several works investigated the outcomes of the voter model by arranging agents in complex networks (e.g.,~\cite{redner01}).
The system can evolve asynchronously or synchronously. In the first case, at each time step, only one agent is considered, whereas in the second case (i.e., the asynchronous one) all the agents change opinion at the same time step.
We implement the voter model by the asynchronous strategy in fermionic networks, in order to evaluate if the system reach the ordered phase (i.e., all the agent share the same opinion) varying the temperature over time. 
In particular, the value of $T$ varies from $T=100$ to $T=60$ (with a temperature-step equal to $1$), and then the system is heated up to $T=100$. Notably, considering the time scale, the time steps corresponding to $T = 100$ are $t = 0$ and $t = 7.7 \cdot 10^{6}$ (i.e., the first and the last time steps, respectively), whereas the temperature $T = 60$ is reached at $t = 4 \cdot 10^{6}$. We recall that the voter model over adaptive networks has already been studied in~\cite{benczik01}.
Figure~\ref{fermionic_network_voter} shows results of numerical simulations. In order to study if the system reaches an ordered phase, we analyze the magnetization of the system over time.
In particular, considering Eq.~\ref{eq:magnetization}, the system reaches an ordered phase (i.e., all agents have the same opinion) when $M \to 1$, whereas the opposite happens if $M \sim 0$ (i.e., the two opinion coexist with equal distribution). 
Notably, it is interesting to observe that these dynamics do not seem affected by the variation of temperature, as the magnetization of the system linearly increases without any critical points (see panel \textbf{a} of Figure~\ref{fermionic_network_voter}).
It is worth to recall that the voter model has been implemented on fermionic networks, varying the temperature in both directions (i.e., cooling and heating the system). 
Moreover, we show that a very similar behavior is obtained also by playing the voter model in a classical random network (with the same number of nodes) ---see panel \textbf{b} of Figure~\ref{fermionic_network_voter}
Since, as shown in Figure~\ref{fermionic_network_voter}, the population reaches an ordered phase (i.e., all agents have the same opinion), we can state that no relation can be identified between the underlying fermionic dynamics of networks and the analyzed process (i.e., the asynchronous voter model). 
\section{Conclusion}\label{sec:conclusions}
The main target of these analysis is to investigate the behavior of fermionic networks, varying the system temperature, in order to evaluate their potential as a computational framework. In particular, fermionic networks can be used to generate adaptive networks, i.e., networks whose structure varies over time.
In general, the structural properties are, as expected, strongly affected by the variation of the temperature, whereas it is interesting to observe that the dynamical process implemented related to opinion dynamics (i.e., the voter model) does not seem to be affected by the temperature.
In particular, the magnetization of the system linearly increases over time up to $1$, hence an ordered phase is achieved after a number of time steps, without that cooling and heating the system affects its increasing tendency.
It is important to observe that, in this kind of models, physical parameters as the temperature require an opportune mapping to a parameter/phenomenon belonging to the considered domain. For instance, considering a network system the temperature can represent the level of competitiveness of the system itself (e.g., a set of web sites that aim to increase their connectivity, or a set of companies that aim to increasing the amount of their customers). 
In order to conclude, we deem that fermionic networks can be considered a useful model to generate adaptive networks and to represent complex systems varying over time, although a preliminary task of mapping be performed between the quantum world and the analyzed domain; in particular, by considering the role of the temperature in the modeled system.
\section*{Acknowledgments}
The author wishes to thank Fondazione Banco di Sardegna for supporting his work.

\end{document}